\documentclass[letterpaper, 10pt, conference]{ieeeconf}

\IEEEoverridecommandlockouts

\usepackage{graphicx,array}
\usepackage{amssymb,amsmath}
\usepackage{cite}
\def\r{\mathbb{R}}

\def\ae{\varkappa}

\title{Estimation of transient process for singularly perturbed synchronization system
with distributed parameters}

\author{ Vera Smirnova$^{1}$, Anton V. Proskurnikov$^{2}$ and Natalia V. Utina$^{3}$
\thanks{*The work was supported by the European Research Council (ERCStG-
307207) and RFBR, grant 12-01-00808. The example (Section V) was supported solely by
Russian Scientific Foundation, grant 14-29-00142.}
\thanks{$^{1}$Vera Smirnova is with the Department of Mathematics,
      St.Petersburg State University of Architecture and Civil Engineering. She is also with
      the Department of Mathematics and Mechanics, St.Petersburg State University,
      St.Petersburg, Russia, {\tt\small root@al2189.spb.edu}}%
\thanks{$^{2}$Anton Proskurnikov is with the Research Institute of Technology and Management, University of Groningen. He is also with
      St.Petersburg State University, ITMO University and Institute for Problems of Mechanical Engineering RAS,
      St.Petersburg, Russia
      {\tt\small avp1982@gmail.com}}%
\thanks{$^{3}$Natalia Utina is with the Department of Mathematics,
      St.Petersburg State University of Architecture and Civil Engineering.
      {\tt\small unv74@mail.ru}}%
}
\begin{document}

\maketitle

\begin{abstract}

Many systems, arising in electrical and electronic engineering 
are based on controlled phase synchronization of several periodic processes (``phase synchronization'' systems, or PSS). Typically such systems are featured by the \emph{gradient-like behavior}, i.e.
the system has infinite sequence of equilibria points, and any solution converges to one of them. This property however says nothing about the transient behavior of the system, whose important qualitative index is the maximal phase error. 
The synchronous regime of gradient-like system may be preceded by cycle slipping, i.e. the increase of the absolute phase error.
Since the cycle slipping is considered to be undesired behavior of PSSs, it is important to find efficient estimates for the number of slipped cycles.
In the present paper, we address the problem of cycle-slipping for phase synchronization systems described by integro-differential Volterra equations with a small  parameter at the higher derivative.
New effective estimates for a number of slipped cycles  are obtained by means of Popov's method of ``a priori integral indices''. The estimates are uniform with respect to the small  parameter.

\end{abstract}


\section{INTRODUCTION}

A vast range of physical and mechanical systems are described by ordinary differential and integro-differential 
equations with a small  parameter at the higher derivative. Such  equations  are usually called singularly perturbed, 
since the order of unperturbed equation is lower than the order of the perturbed one.
So in the absence of special simplifying assumptions, the electron generator is described by a third order differential equation with a small parameter \cite{Caprioli:63},  
Van der Pol equation being the special case when the small parameter is equal to zero. 

The asymptotic properties of singularly perturbed equations may differ from these of unperturbed  ones. So the
problems of stability and oscillations for various singularly perturbed integro-differential 
equations became the subject for special research \cite{Imanaliev}. The armamentarium of singularly perturbed systems is a backbone of the mathematical
of theory of systems with different time scales, arising in different areas of engineering and natural sciences,
from mechanics to mathematical biology \cite{Kokotovic:99}.

In this paper we examine asymptotic behavior of singularly perturbed \emph{phase synchronization systems} (PSS).
The PSSs are based on the principle of \emph{phase synchronization} \cite{Leonov:06}.
These systems, sometimes referred to as synchronous or pendulum-like control systems, involve periodic nonlinearities and typically have infinite sequence of equilibria points.
An important class of PSS is constituted by \emph{phase-locked} systems, which are based on the seminal idea of phase-locked loop (PLL) and widely used in telecommunications and electronics \cite{Margaris}, \cite{Leonov:14-1}. 

Recent decades a vast literature examining asymptotic behavior and other dynamical properties of PSSs has been published,
motivated by numerous applications of these systems in mechanical, electric, electronic and telecommunication engineering.
Most of these papers address the problem of gradient-like behavior, aiming at obtaining conditions which
guarantee convergence of the solutions to equilibria, which means that  the generators of the system are synchronized for any initial state. For details and bibliography see e.g. \cite{Leonov:06}, \cite{Perkin:09-1} and references therein.

But as a rule the synchronous regime of gradient-like system is preceded by cycle slipping, i.e. the increase of the absolute phase error.
Its amplitude depends on the initial state of PSS and is an important  characteristic of the transient process of the system.

The phenomenon of cycle slipping was set forth in the book \cite{Stoker},
for mathematical pendulum with viscous friction proportional
to the square of angular velocity. For mathematical pendulum  the number of full rotations around the point of suspension was called the number of slipped cycles.

The extension of this notion to PSSs is as follows.
Suppose that a gradient-like phase synchronization system has a $\Delta$ - periodic input and
let $\sigma(t)$  be its phase error. They say that the output function $\sigma(t)$ has slipped  $k\in {\bf{N}} \bigcup$ $ \{0\}$  cycles if
there exists such a moment
$\widehat{t} \geq 0 $  that
\begin{equation}
|\sigma(\widehat{t})- \sigma(0)| = k \Delta,
\end{equation}
however for all $t \geq 0 $ one has
\begin{equation}
|\sigma(t)- \sigma(0)| < (k+1)\Delta.
\end{equation}

So to give the adequate description for behavior of PSSs one must establish possibly close estimates for the number of slipped cycles. And since large number of slipped cycles is undesirable for PSSs the problem of its estimation is important.

In the paper \cite{Ershova:83} the problem of cycle slipping was considered  for multidimensional PSSs.
By periodic Lyapunov-like functions and the Kalman-Yakubovich-Popov (KYP) lemma some frequency-algebraic estimates were obtained.
The results of \cite{Ershova:83} were formulated in terms of LMI-solvability in \cite{YangHuang:07}. The estimates of \cite{Ershova:83} were extended to discrete-time and distributed parameter PSSs in the paper \cite{Smirnova:06} and  the monograph\cite{LeonovReitmannSmirnova} respectively. Distributed parameter PSSs were investigated by the method of a priori integral indices with the help of Popov-like functionals of special type.
In paper \cite{Perkin:14-2} the generalized Popov-like functionals from  \cite{Perkin:12}
are exploited for estimation of cycle-slipping for distributed parameter PSSs.

In this paper we make frequency-algebraic estimates from \cite{Perkin:14-2} more exact and extend them for singularly 
perturbed distributed parameter PSSs.

\section{Problem setup}

Consider a distributed parameter synchronization system which is described by the integro-differential Volterra equation
with a small parameter at the higher derivative:
\begin{equation}
 \begin{array}{l}
   \mu \ddot \sigma_\mu (t) + \dot \sigma_\mu (t) = \alpha(t)+ \rho \varphi (\sigma_\mu (t-h))- \\
 -\int \limits_0^{t} \gamma(t-\tau)\varphi((\sigma_\mu (\tau)) \,d\tau \ \ \ (t \ge 0).
                                                                                                                                   \label{1}
\end{array}
\end{equation}
Here ${\mu>0}$, ${h\ge 0}$, ${\rho \in{\bf R}}$, ${\gamma, \alpha : [0,+\infty)\to {\bf R}}$; ${\varphi:{\bf R}\to {\bf R}}$.
The map ${\varphi}$ is assumed ${C^1}$-smooth and ${\Delta}$-periodic with two simple isolated roots on ${[0,\Delta)}$.
The kernel function ${\gamma(\cdot)}$ is piece-wise continuous, the function ${\alpha(\cdot)}$ is continuous.
For each ${\mu}$ the solution of~(\ref{1}) is defined by specifying initial condition
\begin{equation}
  \sigma_\mu(t)|_{t\in [-h,0]} = \sigma^0(t),                                                                                                             \label{2}
\end{equation}
where ${\sigma^0(\cdot)}$ is continuous and ${\sigma(0+0)=\sigma^0(0)}$.

We assume for definiteness that
\begin{equation}
  \int \limits_0^\Delta \varphi(\sigma) \,d\sigma  \le 0.             \label{3}
\end{equation}

We assume also that the linear part of~(\ref{1}) is stable:
\begin{equation} \label{alpha-1}
|\alpha(t)| + |\gamma(t)| \le Me^{-rt} \ \ \ (M,r > 0).
\end{equation}

Let
$$ 
\alpha_1 = \inf\limits_{\sigma \in [0,\Delta)} \frac{d\varphi}{d\sigma}, \ \  \alpha_2 = \sup\limits_{\sigma \in [0,\Delta)} \frac{d\varphi}{d\sigma}.
$$
Then
$$ 
\alpha_1 \le \frac{d\varphi}{d\sigma} \le \alpha_2, \ \ \forall \sigma \in {\bf R}.
$$
Notice that ${\alpha_1<0<\alpha_2}$.

In paper \cite{Pak:13}  
some sufficient frequency-algebraic conditions for gradient-like behavior of equation~(\ref{1}) were obtained,
uniform with respect to the parameter ${\mu}$.
They are formulated in terms of the transform function of the linear part of
the unperturbed Volterra equation $(\mu=0)$
\begin{equation}
 \begin{array}{l}
   \dot \sigma (t) = \alpha(t)+ \rho \varphi (\sigma (t-h)) 
 -\int \limits_0^{t} \gamma(t-\tau)\varphi((\sigma (\tau)) \,d\tau.
                                                                                                                 \label{4}
\end{array}
\end{equation}
The transfer function for (\ref{4}) is defined as follows 
\begin{equation}
   K(p) = -\rho e^{-hp}  + \int \limits_0^{t} \gamma(t) e^{-pt} \,dt \ \ \ (p \in {\bf C}).
                                                                                                                                      \label{5}
\end{equation}

Our goal is to estimate maximal deviations ${ \sup\limits_{t\ge 0} |\sigma_\mu(t) - \sigma_\mu(0)|}$
for a gradient-like system.
Precisely, we are going to obtain estimates for the number of slipped cycles.

\section{Frequency-algebraic estimates in case of the number of cycles slipped for unperturbed equation}

Consider a distributed  parameter system with a periodic nonlinearity, described by the integro-differential equation
(\ref{4}).
This equation is the  unperturbed equation for the equation (\ref{1}). 
Its  transfer function from the input $\varphi$
to the output  $(-\dot{\sigma})$ is defined by (\ref{5}). 
The initial condition for (\ref{4}) has the form
\begin{equation}
\sigma(t)|_{t\in [-h,0]} = \sigma^0(t).                                                                                                             \label{eqn_111}
\end{equation}

Let us obtain certain estimates for the number of slipped cycles for the unperturbed equation.
We start with the following technical lemma which is a cornerstone point in estimating the number of slipped cycles.

 {\bf Lemma 1.} {\it
Suppose there exist such positive
$\vartheta$, $\varepsilon$, $\delta$, $\tau$
that
for all $\omega\geq 0$ the frequency-domain inequality holds:
\begin{equation}\label{eqn_100}
\begin{array}{l}
Re \{ \vartheta K(i\omega)- \tau(K(i\omega)+\alpha_1^{-1}i\omega)^*(K(i\omega)+\alpha_2^{-1}i\omega)\}  \\
- \varepsilon |K(i\omega)|^2-\delta \geq 0 \quad (i^2=-1).
\end{array}
\end{equation}

\noindent
Then the following quadratic functionals
$$
\begin{array}{c}
I_T=\displaystyle\int\limits_0^T \left\{\vartheta\dot{\sigma}(t)
\varphi(\sigma(t))+\varepsilon\dot{\sigma}^2(t) +\delta\varphi^2(\sigma(t))+\right.\\
\left.+\tau(\alpha_1^{-1}\dot{\varphi}(\sigma(t))-\dot{\sigma}(t))
(\alpha_2^{-1}\dot{\varphi}(\sigma(t))-\dot{\sigma}(t))\right\}dt
\end{array}
$$
are uniformly bounded along the solution
of (\ref{4}):
\begin{equation} \label{eqn_101}
I_T\leq Q,
\end{equation}
where $Q$ does not depend on $T$.
}
\begin{proof}
Let $\sigma(t)$ be the solution of
(\ref{4}), (\ref{eqn_111}) and $\eta(t) = \varphi(\sigma(t))$. Let $T$  be
an arbitrary positive number.
If $\eta(0) \ne 0$ we determine
\[
v(t) = \left\{
                   \begin{array}{ccl}
                     0 & \text{for} & t<0, \\
                     t & \text{for} & t\in [0,1], \\
                     1 & \text{for} & t>1 \\
                   \end{array}
\right\},
\]
in case $\eta(0)= 0$ we put $v(t)=1$.

We introduce auxiliary functions
\begin{eqnarray*}
  \sigma_0(t):=\alpha(t) + (1-v(t-h))\rho\eta(t-h)- \\
              - \int\limits_0^t
              (1-v(\tau))\gamma(t-\tau)\eta(\tau)d\tau,
\end{eqnarray*}
\[
\zeta_T(t):= \left\{
                   \begin{array}{cl}
                     \eta(t) & t\leq T, \\
                     \eta(T)e^{\lambda(T-t)} & t>T \quad (\lambda>0)
                   \end{array}
\right\},
\]
\[
\eta_T(t):= v(t)\zeta_T(t),
\]
\[
\sigma_T(t): = \rho \eta_T(t-h) -\int\limits_0^t
\gamma(t-\tau)\eta_T(\tau)d\tau.
\]
 For $t\in[0,T]$ we have
\begin{equation}\label{eqno_132}
\dot{\sigma}(t)=\sigma_0(t) +\sigma_T(t).
\end{equation}

For any $T>0$, consider a functional
$$
\begin{array}{c}
\rho_T:=\displaystyle\int\limits_0^{+\infty} \left\{\vartheta\sigma_T(t)
\eta_T(t)+\delta\eta_T^2(t)+\varepsilon\sigma^2_T(t)+\right. \\
\left.+\tau(\sigma_T(t)-\alpha_1^{-1}\dot{\eta}_T(t))
(\sigma_T(t)-\alpha_2^{-1}\dot{\eta}_T(t))\right\}dt.
\end{array}
$$

It is demonstrated in \cite{LeonovReitmannSmirnova} by means of the Popov method of ``a priori integral indices'' \cite{Rasvan:06} 
that the frequency-domain inequality (\ref{eqn_100}) implies
\begin{equation} \label{eqn_34}
\rho_T\leq 0\quad\forall T>0.
\end{equation}

It can be easily shown that
\begin{equation} \label{eqn_94}
\rho_T = I_T+I_{1T}+I_{2T}+I_{4T},
\end{equation}
where the addends $I_{1T},I_{2T},I_{4T}$ are defined by
\begin{eqnarray*}
&&I_{1T}: = -\int\limits_0^{\ell}\left\{
\vartheta(1-v(t))\dot{\sigma}(t)\eta(t)+\delta(1-\mu^2)\eta^2(t)+\right.
\\&&
\left. +\alpha_1^{-1}\alpha_2^{-1}\tau\dot{\eta}^2(t)-
\left(\dot{\widehat{v(t)\eta(t)}}\right)^2\alpha_1^{-1}\alpha_2^{-1}\tau+ \right.
\\&&
\left. +\tau\left(\alpha_1^{-1}+\alpha_2^{-1}\right)\dot{\sigma}(t)
\left(\dot{\eta}(t)-\dot{\widehat{v(t)\eta(t)}}\right)
\right\}dt,
\end{eqnarray*}
with $\ell=T$, if $T<1$, and $\ell=1$, if $T\geq 1$;
\begin{eqnarray*}
&&I_{2T} := \int\limits_0^{T}\left\{
-\vartheta\sigma_0(t)\eta(t)v(t)-2(\varepsilon+\tau)\dot{\sigma}(t)\sigma_0(t)+
\right.
\\&&
+\left.(\tau+\varepsilon)\dot{\sigma}_0^2(t)+\tau\left(\alpha_1^{-1}+
\alpha_2^{-1}\right)\sigma_0(t)\dot\eta(t)\right\}dt;
\end{eqnarray*}
\begin{eqnarray*}
&&I_{4T} := \int\limits_T^{\infty}\left\{
\vartheta\sigma_T(t)\eta_T(t)+\delta\eta_T^2(t)+(\varepsilon+\tau)\sigma^2_T(t)
-\right.
\\&&
-\left.\left(\alpha_1^{-1}+\alpha_2^{-1}\right)\tau\sigma_T(t)\dot{\eta}_T(t)+\tau\alpha_1^{-1}\alpha_2^{-1}\dot{\eta}^2_T(t)\right\}dt.
\end{eqnarray*}

From (\ref{eqn_34}) and (\ref{eqn_94}) it is immediate that
\begin{equation} \label{eqn_36}
I_T \leq -I_{1T}-I_{2T}-I_{4T}.
\end{equation}

As can be seen from (\ref{alpha-1}), the functionals $I_{1T}$ and $I_{2T}$ are uniformly bounded:
\begin{equation} \label{eqn_37}
|I_{1T}| \leq q_1(\delta,\vartheta,\tau,\alpha_1,\alpha_2),
\end{equation}
\begin{equation} \label{eqn_38}
|I_{2T}| \leq q_2(\vartheta,\varepsilon,\tau,\alpha_1,\alpha_2),
\end{equation}
where $q_1$ and $q_2$ do not depend on $T$. By definition of $\eta_{T}(t)$, 
$$
\begin{array}{l}
I_{4T} = \displaystyle\int\limits_T^{\infty}\left\{\vartheta\sigma_T(t)\eta(T)e^{\lambda(T-t)}+\delta\eta^2(T)e^{2\lambda(T-t)}+\right.\\
\left.+(\varepsilon+\tau)\sigma^2_T(t)+\lambda
\left(\alpha_1^{-1}+\alpha_2^{-1}\right)\tau\sigma_T(t)\eta(T)e^{\lambda(T-t)}+\right.\\
\\\left.+\lambda^2\tau\alpha_1^{-1}\alpha_2^{-1}\eta^2(T)e^{2\lambda(T-t)}\right\}dt.
\end{array}
$$
Introducing the constant
\begin{equation} \label{eqn_39}
\lambda :=
\sqrt{\displaystyle\frac{\delta|\alpha_1|\alpha_2}{\tau}},
\end{equation}
it can be shown that
\begin{equation}
\begin{array}{c}
  I_{4T}\geq
-\displaystyle\int\limits_T^{\infty}\displaystyle\frac{W^2}{4(\varepsilon+\tau)}
\cdot  \eta^2(T)e^{2\lambda(T-t)}dt\nonumber,
\end{array}
\end{equation}
where
$
W = \vartheta+\sqrt{\tau\delta|\alpha_1|\alpha_2}\left(\alpha_1^{-1}+\alpha_2^{-1}\right),
$
and hence
\begin{equation} \label{eqn_41}
I_{4T} \geq -q_3(\delta,\vartheta,\tau,\alpha_1,\alpha_2)=\frac{\sqrt{\tau}\,W^2\max{\varphi^2(\sigma)}}{8(\varepsilon+\tau)\sqrt{\delta|\alpha_1|\alpha_2}}.
\end{equation}
The inequalities (\ref{eqn_36}), (\ref{eqn_37}), (\ref{eqn_38}), and (\ref{eqn_41}) yield that
\begin{equation} \label{eqn_42}
I_T\leq q_1 + q_2 + q_3 \equiv Q(\vartheta, \delta, \varepsilon, \tau,
\alpha_1, \alpha_2),
\end{equation}
which finishes the proof of Lemma 1.
\end{proof}

The value of $Q$ from (\ref{eqn_101}) 
may be found explicitly.

{\bf Lemma 2. }
{\it 
Let $|\alpha_1|=\alpha_2$ and $\varphi(\sigma(0))=\varphi(\sigma(T))=0$. 
Then if the conditions of Lemma 1 are fulfilled the following estimate holds 
\begin{equation} \label{lemma2-1}
I_T\leq q:=\frac{1}{r}(\vartheta Mm+2(\varepsilon+\tau)Mm(\frac{M}{r}+\rho)+(\varepsilon+\tau)\frac{M^2}{2}),
\end{equation}
where 
$$
m=\sup{\varphi(\sigma)}.
$$
}

\begin{proof}
Let us exploit Lemma 1 for $\varphi(\sigma(0))=0$ and $\varphi(\sigma(T))=0$.
In this case $I_{1T} = 0$ and $I_{4T} \geq 0$.
So it follows from (\ref{eqn_36})  that

\begin{equation} \label{eqno_300}
\begin{array}{c}
I_T \leq  -I_{2T}.
\end{array}
\end{equation}

Since $|\alpha_1|=\alpha_2$
\begin{equation} \label{508}
\begin{array}{c}
I_T\leq\displaystyle\int\limits_0^\infty |\vartheta\alpha(t)
\varphi(\sigma(t))+(\tau+\varepsilon)\alpha^2(t) +
2(\tau+\varepsilon)\alpha(t)\cdot \\
\cdot(\rho\varphi(\sigma(t-\tau))-
2(\tau+\varepsilon)\displaystyle\int\limits_0^t\gamma(t-\tau)\varphi(\sigma(\tau))d\tau)|dt
\end{array}
\end{equation}
Hence it follows that the estimate  (\ref{lemma2-1}) is true.

\end{proof}


The next two theorems give estimates for the number of slipped cycles. To start with, we introduce auxiliary functions
$$
\Phi(\sigma) =
\sqrt{(1-\alpha_1^{-1}\varphi'(\sigma))(1-\alpha_2^{-1}\varphi'(\sigma))},
$$
$$
P(\varepsilon, \tau,\sigma) = \sqrt{\varepsilon +\tau\Phi^2(\sigma)},
$$
$$
r_{1j}(k, \vartheta, \varepsilon, \tau, x) =
\frac{\int\limits_0^\Delta\varphi(\sigma)d\sigma +
(-1)^j\frac{x}{\vartheta k}}{\int\limits_0^\Delta
|\varphi(\sigma)| P(\varepsilon, \tau,\sigma) d\sigma} \quad (j=1,2)
$$
$$
Y_j(\sigma) =
\varphi(\sigma)- r_{1j}|\varphi(\sigma)|P(\varepsilon, \tau,\sigma) \quad  (j=1,2).
$$

{\bf Theorem 1.} {\it
Suppose there exist such positive
$\vartheta$, $\varepsilon$, $\delta$, $\tau$ and
natural $k$   that the following conditions are fulfilled:

1) for all $\omega\geq 0$ the frequency-domain inequality (\ref{eqn_100})
 holds;

2)
\begin{equation} \label{eqn_5}
4\delta > \vartheta^2(r_{1j}(k,\vartheta,\varepsilon, \tau, Q))^2  \quad  (j=1,2),
\end{equation}
where $Q$ is given by (\ref{eqn_101}). Then any solution of (\ref{4}) slips less than $k$ cycles, that is, the inequalities hold
\begin{equation} \label{eqn_33}
|\sigma(0)-\sigma(t)|<k\Delta\quad\forall t\ge 0.
\end{equation}
}

The proof of Theorem 1 is presented in \cite{Perkin:14-2}.

To proceed with the next result, we introduce the following functions for $j=1,2$
$$
r_j(k, \vartheta, x) := \frac{\int\limits_0^\Delta\varphi(\sigma)d\sigma
+ (-1)^j\frac{x}{\vartheta k}}{\int\limits_0^\Delta
|\varphi(\sigma)|d\sigma},
$$
$$
r_{0j}(k, \vartheta, x) :=
\frac{\int\limits_0^\Delta\varphi(\sigma)d\sigma +
(-1)^j\frac{x}{\vartheta k}}{\int\limits_0^\Delta \Phi(\sigma)
|\varphi(\sigma)|d\sigma},
$$
$$
F_j(\sigma) =\varphi(\sigma) - r_j|\varphi(\sigma)|,
$$
$$
\Psi_{j}(\sigma) = \varphi(\sigma) -
r_{0j}|\varphi(\sigma)|\Phi(\sigma)
$$
and matrices
$ T_j(k, \vartheta, x):=$
\[
:=\left\|
\begin{array}{ccccc}
\varepsilon & \displaystyle\frac{a\vartheta r_j(k,\vartheta,x)}{2} & 0 \\
\displaystyle\frac{a\vartheta r_j(k,\vartheta,x)}{2} &  \delta &
\displaystyle\frac{a_0\vartheta r_{0j}(k,\vartheta,x)}{2} \\
0 &  \displaystyle\frac{a_0\vartheta r_{0j}(k,\vartheta,x)}{2} & \tau
\end{array}
\right\|,
\]
where $a\in[0,1]$ and $a_0:=1-a$.

 {\bf Theorem 2.} {\it
Suppose there exist positive  
$\vartheta$, $\varepsilon$, $\delta$, $\tau$, $a \in [0,1] $ and
natural $k$ satisfying the conditions as follows:

1) for all $\omega\geq 0$ the frequency-domain inequality (\ref{eqn_100}) holds;

2) the matrices $T_j(k,\vartheta,Q)$ $(j=1,2)$  where the value of $Q$
is defined by (\ref{eqn_101}),
are positive definite.

\noindent
Then for the solution of (\ref{4}) the inequality (\ref{eqn_33}) holds.
}

\begin{proof}
Let $\sigma(t)$ be the solution of
(\ref{4}), (\ref{eqn_111}).

Let $\varepsilon_0>0$ be so small that matrices
$T_j(Q+\varepsilon_0)$ are positive definite.
We consider the functions $F_j(\sigma)$, $\Psi_j(\sigma)$ \, $(j=1,2)$ 
with
$$
r_j=r_j(k,\vartheta,Q+\varepsilon_0),
$$
$$
r_{0j}=r_{0j}(k,\vartheta,Q+\varepsilon_0)
$$
and the functionals $I_T$ from Lemma 1.

It is true that
\begin{equation} \label{eqn_43}
\begin{array}{c}
  I_T = \vartheta a \int\limits_{\sigma(0)}^{\sigma(T)} F_j(\sigma)
  d\sigma + \vartheta a_0 \int\limits_{\sigma(0)}^{\sigma(T)}
  \Psi_j(\sigma)+\\+
  \int\limits_{0}^{T} \left\{
  \vartheta\dot\sigma(t)\varphi(\sigma(t))+\varepsilon\dot\sigma^2(t)+\delta\varphi^2(\sigma(t))-\right.\\
\left.-\vartheta a F_j(\sigma(t))\dot\sigma(t) - \vartheta a_0 \Psi_j(\sigma(t))\dot\sigma(t) +\right.\\
 \\\left.+\tau\dot\sigma^2(t)\Phi(\sigma(t))\right\} dt \quad (j=1,2).
\end{array}
\end{equation}

In virtue of condition 2) of the theorem, the third term in the right hand part of (\ref{eqn_43})
is the integral of positive definite quadratic form.
So
\begin{equation}
\begin{array}{r}
I_T \geq \vartheta\left(a\int_{\sigma(0)}^{\sigma(T)}F_j(\sigma)d\sigma
+a_0\int_{\sigma(0)}^{\sigma(T)}\Psi_j(\sigma)d\sigma \right)  \\
(j=1,2).
\end{array}
\end{equation}

Suppose that
$$
\sigma(t_1)=\sigma(0)+k\Delta.
$$
Then
$$
\int_{\sigma(0)}^{\sigma(t_1)}F_1(\sigma)d\sigma=k
\int_{0}^{\Delta}F_1(\sigma)d\sigma=\displaystyle\frac{1}{\vartheta}(Q+\varepsilon_0),
$$
$$
\int_{\sigma(0)}^{\sigma(t_1)}\Psi_1(\sigma)d\sigma=k
\int_{0}^{\Delta}\Psi_1(\sigma)d\sigma=\displaystyle\frac{1}{\vartheta}(Q+\varepsilon_0).
$$
Then
\begin{equation} \label{eqn_44}
I_{t_1} \geq Q +\varepsilon_0 > Q.
\end{equation}
which contradicts with (\ref{eqn_42}). So our hypothesis is wrong.
With the help of $F_2(\sigma)$ and $\Psi_2(\sigma)$ we prove that
$$
\sigma(t)\neq\sigma(0)-k\Delta.
$$

As a result for all $t>0$
$$
\sigma(0)-k\Delta < \sigma(t) < \sigma(0)+k\Delta.
$$

Theorem 2 is proved.
\end{proof}


{\bf Theorem 3} {\it
Let $|\alpha_1|=\alpha_2$ and $\sigma(0)=\sigma_0$ where $\varphi(\sigma_0)=0$.
Suppose there exist such positive
$\vartheta$, $\varepsilon$, $\delta$, $\tau$, $a \in [0,1] $ and
natural $k$   that the following conditions are fulfilled:

1) for all $\omega\geq 0$ the frequency-domain inequality(\ref{eqn_100}) holds;

2) the matrices $T_j(k,\vartheta,q)$ $(j=1,2)$, with $q$ defined by (\ref{lemma2-1}),  
are positive definite.

\noindent
Then for any solution of (\ref{4}) the estimate (\ref{eqn_33}) holds.
}

\begin{proof}
Let us repeat the proof of Theorem 2 with 
$$
r_j=r_j(k,\vartheta,q+\varepsilon_0),
$$
$$
r_{0j}=r_{0j}(k,\vartheta,q+\varepsilon_0).
$$
Then
\begin{equation} \label{theorem3-1}
I_{t_j} > q \quad (j=1,2).
\end{equation}
Since $\varphi(\sigma(0))=0$ we conclude that   $\varphi(\sigma(t_1))=0$ and $\varphi(\sigma(t_2))=0$.
Then it follows from Lemma 2 that 
\begin{equation*} 
I_{t_j} \leq q \quad (j=1,2),
\end{equation*}
which contradicts (\ref{theorem3-1}).

\end{proof}

\section{Example}
Let us consider a phase-locked loop (PLL) with a proportional integral lowpass filter, a sine-shaped characteristic of phase frequency detector and a time-delay in the loop.
Its mathematical description is borrowed from \cite{Belustina}:
\begin{equation} \label{eqno_307}
\ddot\sigma(t)+\frac{1}{T}\dot\sigma(t)+\varphi(\sigma(t-h))+sT\dot\varphi(\sigma(t-h))=0,
\end{equation}
$$
\varphi(\sigma)=\sin\sigma -\beta, \, s \in (0,1), \,  \beta \in (0,1], \, h>0, \, T>0.
$$
The differential equation (\ref{eqno_307}) can be reduced to integro-differential equation  (\ref{4})
with
\begin{equation*}
\gamma(t) = \left\{
                   \begin{array}{ccl}
                     0, & t<h, \\
                    (1-s)e^{-\frac{t-h}{T}}, & t\geq h \\
                   \end{array}
\right\},
\end{equation*}
\begin{equation*}
\alpha(t) =e^{-\frac{t}{T}}(b-(1-s)J),
\end{equation*}
where $b=\dot\sigma(0)+sT\varphi(\sigma(-h))$ and
\begin{equation*}
J =
\left
\{
\begin{array}{ccl}
      \displaystyle
\int\limits_{-h}^{t-h}e^{\frac{\lambda+h}{T}}\varphi(\sigma(\lambda))d\lambda, & t\leq h, \\
       \displaystyle
\int\limits_{-h}^{0}e^{\frac{\lambda+h}{T}}\varphi(\sigma(\lambda))d\lambda, & t> h
\end{array}
\right\}.
\end{equation*}

The transfer function of the lowpass filter here has the form:
\begin{equation*}
K(p)=T\frac{Tsp+1}{Tp+1}e^{-ph}
\end{equation*}

We suppose that $\varphi(\sigma(0))=0$ and apply Theorem 3.

Let $\alpha_2=-\alpha_1=1,\,\vartheta=1, \, a=1$.
The assumption 1) of  Theorem 3 shapes into
\begin{equation} \label{eqno_308}
\begin{array}{ll}
\Omega(\omega)\equiv \tau T^2\omega^4 + \omega^2(T^3s\cos{\omega h}-T^4s^2(\varepsilon+\tau) +\\
+\tau-\delta T^2) -T^2(1-s)\omega\sin{\omega h} +T\cos{\omega h} - \\
-(\varepsilon+\tau)T^2-\delta \geq 0 \quad  \forall \omega;
\end{array}
\end{equation}
whereas condition 2) may be rewritten as
\begin{equation} \label{eqno_309}
2\sqrt{\varepsilon\delta} > \frac{2\pi\beta+q_2k^{-1}}{4(\beta\arcsin\beta+\sqrt{1-\beta^2})}.
\end{equation}

Notice that for all $\omega\in\r$ one has
\begin{equation*}
\begin{array}{ll}
\Omega(\omega)\geq \Omega_0(\omega)\equiv (\tau T^2-\frac{1}{2}T^3sh^2)\omega^4 + (T^3s-\\
-T^4s^2(\varepsilon+\tau) +\tau-\delta T^2-\frac{1}{2}Th^2 -(1-s)T^2h)\omega^2+\\
+(T-(\varepsilon+\tau)T^2-\delta), \quad \forall \omega
\end{array}
\end{equation*}
and $\Omega(\omega)\approx\Omega_0(\omega)$ when $\omega h<<1$.

We consider the case $T\leq 0.9,\, h_0= \frac{h}{T}\leq 1$, since for small $T$ and small $h$ the PLL is
gradient-like for all $\beta \in (0,1]$ \cite{Belustina}.
Let us choose $\varepsilon=\frac{\beta_0}{T},\, \delta=\alpha_0T,\, \tau=\gamma_0 T^3$.
As $\Omega(0)=\Omega_0(0)$ it is necessary that $\alpha_0+\beta_0+\gamma_0T^4\leq 1$.
Then the optimal values for $\alpha_0$ and $\beta_0$ are $\alpha_0=\beta_0=\frac{1}{2}(1-\gamma_0T^4)$,
whence $2\sqrt{\varepsilon\delta}=1-\gamma_0T^4$.
For $\gamma_0=\max{\{\frac{1}{2}sh_0^2,\, \frac{1}{2}(h_0+1-s)^2\}}$ the polynomial   $\Omega_0(\omega)$
is nonnegative, $\forall\omega$.
It follows from (\ref{eqno_309}) that the number $k_0$ of cycles slipped satisfies the inequality
\begin{equation*}
k_0\leq r_0:=\lfloor q_2(8\sqrt{\varepsilon\delta}(\beta\arcsin\beta+\sqrt{1-\beta^2})-2\pi\beta)^{-1}\rfloor,
\end{equation*}
where $\lfloor x\rfloor$ stands for the integer floor of $x$.

Let us consider the PLL with $b=K(0)\beta$ \cite{Ershova:83}.
Then by estimating the functional $I_T$ we conclude that the value of $q$ can be defined by the
formula
\begin{equation} \label{eqno_310}
q=T^2(A+Bh_0+Ch_0^2),
\end{equation}
where
\begin{equation} \label{eqno_311}
\begin{array}{ll}
A=(\frac{7}{2}\beta^2 +3), \\
B=3(1-s)(1+\beta)(3\beta+1),\\
C=\frac{3}{2}(1-s)^2(1+\beta)^2.
\end{array}
\end{equation}
It follows from (\ref{eqno_310}), (\ref{eqno_311}) that the number of slipped cycles increases together
with $T$, with $\beta$ or with $h_0$.
Let for example $h_0=1,\, s=0.4,\,T=0.1$.
Then $r_0=1$ for $\beta=0.9$,\, $r_0=2$ for $\beta=0.92$, and $r_0=5$ for $\beta=0.95$.

\section{Estimates for the number of cycles slipped for singularly perturbed equation}

Equation (\ref{1}) can be reduced to integro-differential Volterra equation
\begin{equation} \label{eqn_51}
\begin{array}{l}
\dot{\sigma_\mu}(t)=\alpha_\mu(t)
-\displaystyle\int\limits_0^t\gamma_\mu(t-\tau)\varphi(\sigma_\mu(\tau))d\tau
\quad (t>0),
\end{array}
\end{equation}
where
\begin{equation} \label{eqn_52}
\begin{array}{l}
\alpha_\mu(t)=\displaystyle\dot{\sigma}(0)e^{\frac{-t}{\mu}}+\frac{1}{\mu}
\int\limits_{0}^{t}e^{\frac{\lambda-t}{\mu}}\alpha(\lambda)d\lambda +
\frac{\rho}{\mu}J_0,
\end{array}
\end{equation}
\begin{equation*}
J_0 =
\left
\{
\begin{array}{ccl}
      \displaystyle
\int\limits_{-h}^{t-h}e^{\frac{\lambda+h-t}{\mu}}\varphi(\sigma(\lambda))d\lambda, & t\leq h, \\
       \displaystyle
\int\limits_{-h}^{0}e^{\frac{\lambda+h-t}{\mu}}\varphi(\sigma(\lambda))d\lambda, & t> h,
\end{array}
\right\}.
\end{equation*}
\begin{equation} \label{eqn_53}
\gamma_\mu(t)=\displaystyle\frac{1}{\mu}
\int\limits_{0}^{t}e^{\frac{\lambda-t}{\mu}}\gamma(\lambda)d\lambda -
\frac{\rho}{\mu}
\left
\{
\begin{array}{ccl}
e^{\frac{h-t}{\mu}}, &  t\geq h,\\
0 , & t< h
\end{array}
\right \}. 
\end{equation}

The transfer function for equation (\ref{eqn_51}) is as follows 
\begin{equation} \label{eqn_54}
K_\mu(p)=\frac{K(p)}{1+\mu p}.
\end{equation}
Let 
$$
q_0=q+(\vartheta m+2(\varepsilon+\tau)m(\frac{M}{r}+\rho))\rho mh +(\varepsilon+\tau) \rho^2 m^2 h,
$$
where $q$ is defined by (\ref{lemma2-1}).

{\bf Theorem 4} {\it
Let $\alpha_2=|\alpha_1|$ and
$\sigma(0)=\sigma_0$ where $\varphi(\sigma_0)=0$.
Suppose there exist such positive
$\vartheta$, $\varepsilon$, $\delta$, $\tau$, $a \in [0,1] $ and
natural $k$   that the following conditions are fulfilled:

1) for all $\omega\geq 0$ the frequency-domain inequality(\ref{eqn_100}) holds;

2) the matrices $T_j(k,\vartheta,q_0)$ $(j=1,2)$  
are positive definite.

\noindent
Then there exists such value $\mu_0$ that  for all $\mu \in (0,\mu_0)$ the following assertion is true: 
for any solution of (\ref{4}) the estimates 
\begin{equation} \label{Theorem4-1}
|\sigma_\mu(0)-\sigma_\mu(t)|<k\Delta\quad\forall t\ge 0
\end{equation}
hold.
}
\begin{proof}
Let $|\alpha_1|=\alpha_2=\ae$.
In our case the inequality (\ref{eqn_100}) takes the form
\begin{equation}\label{Theorem4-2}
\Pi(\omega)\geq 0,
\end{equation}
where
\begin{equation}\label{503}
\Pi(\omega):=\tau\ae^{-2}\omega^2 +
\vartheta Re \{K(i\omega)\}- (\varepsilon+\tau)|(K(i\omega)|^2-\delta.
\end{equation}
Consider $\bar\delta<\delta$ and $\bar\Pi(\omega)=\Pi(\omega)+\delta-\bar\delta $. 
We have 
\begin{equation}\label{Theorem4-3}
\bar\Pi(\omega)> 0 \quad \forall{\omega}.
\end{equation}
Let us substitute in matrices $T_j(k,\vartheta,q_0)$ the value $\delta$ by $\bar\delta$ and denote 
the new matrices by $\bar T_j(k,\vartheta,q_0)$
Let $\delta-\bar\delta$ be so small that $\bar T_j(k,\vartheta,q_0)$ are positive definite.

For integro-differential Volterra equation (\ref{eqn_51}) one can apply Theorem 3.
For transfer function (\ref{eqn_54}) the frequency-domain inequality (\ref{eqn_100}) 
with $\vartheta$, $\varepsilon$, $\bar\delta$, $\tau$ takes the form 
\begin{equation}\label{500}
\begin{array}{r}
\Pi_\mu(\omega):=
Re \{ \vartheta K_\mu(i\omega)\}- (\varepsilon+\tau)|(K_\mu(i\omega)|^2+\\
+\tau\ae^{-2}\omega^2-\bar\delta \geq 0.
\end{array}
\end{equation}
or
\begin{equation}\label{501}
\Pi(\omega)+
\begin{array}{c}
\vartheta \mu\omega Im\{K(i\omega)\}+ \tau\mu^2\ae^{-2}\omega^4 - \bar\delta\mu^2\omega^2 \geq 0.
\end{array}
\end{equation}

Introduce a constant 
\begin{equation}\label{504}
\Omega :=\ae\sqrt{\frac{\bar\delta}{\tau}}.
\end{equation}
Let us fix a certain $\tilde\mu>0$.
Since the value of $|K(i\omega)|$ is bounded for $\omega\in{\bf R}$ we can assert
that there exists such number $\Omega_0>\Omega$ that the inequality (\ref{501})
is true for $\omega>\Omega_0,\,\mu\leq\tilde\mu$.
Let
\begin{equation}\label{505}
\begin{array}{l}
\delta_1= \inf\limits_{\omega \in [0,\Omega_0]} \bar\Pi(\omega) , \\  
L_1 = 2\sup\limits_{\omega \in [0,\Omega_0]}|\vartheta\omega Im\{K(i\omega)\}|. 
\end{array}
\end{equation}
Then if
\begin{equation}\label{505}
\bar\mu<\min\left\{ \frac{\delta_1}{L_1},\,\sqrt{\frac{2\delta_1\tau}{\ae^2\bar\delta}},\,\tilde\mu\right\},
\end{equation}
the inequality (\ref{501}) is true for any $\mu<\bar\mu$ and all $\omega\geq0$.
 
Consider now the functional $I_T$ from Lemma 1 for the solution $\sigma_\mu(t)$ 
of equation (\ref{eqn_51}) and the nonlinear function $\varphi(\sigma_\mu(t))$.

The estimate (\ref{508}) for nonlinear function $\varphi(\sigma_\mu(t))$
and the term $\alpha_\mu(t))$ takes  then the form
$$
\begin{array}{c}
I_T\leq\displaystyle\int\limits_0^\infty \big|\vartheta\alpha_\mu(t)
\varphi(\sigma_\mu(t))+(\tau+\varepsilon)\alpha_\mu^2(t) +\\
+2(\tau+\varepsilon)\displaystyle\int\limits_0^t \gamma_\mu(t-\tau)\varphi(\sigma_\mu(\tau)d\tau)\big|dt
\end{array}
$$
Hence 
\begin{equation}\label{506}
I_T\leq q_\mu, 
\end{equation}
where
\begin{equation}\label{509}
\begin{array}{r}
q_\mu =\displaystyle(\vartheta m+2(\varepsilon+\tau)m(\rho+\frac{M}{r}))(\mu|\dot\sigma(0)|+\frac{M}{r}+\rho mh)+\\
+\displaystyle(\varepsilon+\tau)(\frac{\mu}{2}\dot\sigma^2(0))+\frac{M^2}{2(1-r\mu)^2}(\mu-\frac{4\mu}{1-r\mu}+\displaystyle\frac{1}{r})+\\
+\displaystyle\rho^2 m^2(h+\mu e^{-\frac{h}{\mu}}-\mu)).
\end{array}
\end{equation}
It is clear that
\begin{equation}\label{507}
\lim{q_\mu}= q_0
\end{equation}
as $\mu \rightarrow 0$.

Since matrices $\bar T_j(k,\vartheta,q_0)$ are positive definite
there exists a value $\hat\mu$ small enough that for $\mu\leq\hat\mu$
matrices $\bar T_j(k,\vartheta,q_\mu)$ are positive definite.
Thus if $\mu<\mu_0:=\min\{\bar\mu,\,\hat\mu\}$ the frequency inequality of Theorem 3
for the transfer function $K_\mu(p)$ and the algebraic restrictions on the varying parameters  
are fulfilled.
So for $\sigma_\mu(t)$ the estimate (\ref{Theorem4-1}) is true.
\end{proof}

\section{CONCLUSION}

The paper is devoted to the problem of cycle-slipping for singularly perturbed distributed parameter phase synchronization systems.
The PSSs  described by integro-differential Volterra equations with a small parameter at the higher derivative are addressed. 
The case of differentiable nonlinearities is considered. The problem is investigated with the help of the method of a priori integral indices. In the paper new effective multiparametric frequency-algebraic estimates for the number of slipped cycles  of the output of the system are established. The estimates obtained are uniform with respect to a small parameter.

\bibliographystyle{plain}
\bibliography{literature}

\end{document}